# Nanometric laser trapping based on nanostructured substrates

A. R. Sidorov[1,2], Y. Zhang[1], A. N. Grigorenko[1*], and M. R. Dickinson[1]

[1] *School of Physics and Astronomy, University of Manchester, Manchester, M13 9PL, UK,*

[2] *General Physics Institute, 38 Vavilov str., Moscow 117942, Russia.*

**Abstract:**

Laser trapping near the surface of a nanostructured substrate is demonstrated. Stable microbubbles with radii of 1-20 µm have been created and manipulated with sub-micron precision by a focused laser beam in an immersion oil covering arrays of pairs of gold nanopillars deposited on a glass substrate. The threshold for bubble creation and trapping characteristics depended on near-field coupling of nanopillars. The nanometric laser tweezers showed giant trapping efficiency of $Q \sim 50$ for the trapped microbubbles.

*corresponding author: e-mail: sasha@man.ac.uk

Optical tweezers formed by strongly focused laser beams and introduced by Ashkin *et al.* [1] have been extensively used for trapping and manipulating small-size particles and biological objects [2]. Conventional optical tweezers trap small objects near the focus of a laser beam. As a result, the trapping volume of the conventional tweezers is diffraction limited. It was recently suggested that the trapping volume could be reduced beyond diffraction limit using optical near fields [3]. The nanometric tweezers proposed in [3] rely on strongly enhanced electric fields near the tip of a metallic nanoparticle and offer a smaller trapping volume, a larger trapping force and nanometric precision in object positioning.

In this Letter we report the first experimental realisation of nanometric laser tweezers. The sub-wavelength trapping has been achieved near the surface of a 2D array of gold nanoparticles produced by high-resolution electron beam lithography on a glass substrate. Instead of a single sharply pointed pin suggested in [3], we make use of "nanomolecules" formed by gold nanoparticles arranged in tightly spaced pairs. Such geometry provides excellent control over the critical feature (the gap in the pair) and the frequencies of plasmonic resonances [4,5] important for laser trapping. To demonstrate the action of nanometric laser tweezers we have created, trapped and manipulated microbubbles at the surface of the nanostructured substrate submerged into an immersion oil using a focused laser beam. We found that laser tweezers based on the nanostructured substrate provides a strong trapping force of 1nN for 50mW of laser power, giant trapping efficiency $Q$~50 and a sub-wavelength size of the trap <100nm for the studied microbubbles (for neodymium laser with wavelength 1064nm). Being a simple addition to conventional laser tweezers, the proposed nanostructured substrates could dramatically improve trapping characteristics and could be applied for manipulation and selective analysis of bio and chemical objects.

Figure 1 shows the general scheme of our installation described in detail in Ref. [6]. The trap was created using a 100mW, 1064nm neodymium-doped $YVO_4$ diode pumped solid-state laser, collimated to a beam diameter of 5mm and focused through an oil immersion microscope objective of numerical aperture 1.47 onto the sample substrate. The laser beam power was monitored using a power meter and calibrated to yield the beam power at the sample. In addition to laser light, the sample was illuminated by white light so that the image of the trapped objects could be captured by the CCD camera and transferred to the PC. The studied structures were placed on a *x-y* positioner connected to a rough micrometer translator (with a position accuracy of 1μm) and a fine piezoelectric translator (with a position accuracy of 5nm).

Nanometric laser trapping was realised in our experiments near regular arrays of gold nanopillars fabricated by high-resolution electron-beam lithography on a glass substrate and grouped in tightly spaced pairs. The structures typically covered an area of ≈0.1mm$^2$ and contained ≈10$^6$ pillars. Heights *h* of gold pillars and their diameters were chosen through numerical simulations so that the plasmon resonance of an individual pillar appeared at red-light wavelengths. A number of different structures were studied with *d* between 80 and 160nm and the pair separations *s* between the centres of adjacent pillars in the range 140 to 200nm and, i.e. the gap $\delta=s-d$ between the neighbouring pillars varied from 100nm down to almost zero (overlapping pillars). The data described in this Letter were obtained on three samples with the same pair separation *s*=140nm, the lattice constant *a*=400nm and *h*=90nm, but different diameters of the pillars (and hence different near-field coupling between nanopillars). Figure 2(a) shows an electron micrograph of one of our samples.

At such small pillar separations, the electromagnetic interaction between nanopillars [4,5,7] splits the plasmon resonance of an individual pillar into two resonances for the pillar pair and plasmonic modes of the double pillar "nano-molecule" can be characterised by their parity [7-9]. The antisymmetric resonances were observed at "green" wavelengths while the symmetric resonances were observed in the "red" part of the spectrum, see the curve 1 of Fig. 2(b) depicting the reflection spectrum of the sample measured in the air for normal incident light of TM polarization (the electric field vector directed along the line connecting the pillars in the pair). The spectral positions of plasmon resonance peaks of the double-pillar nano-molecule are determined by the plasmon resonance of the single pillar, the pillar gap and the refractive index of the environment [7]. When the sample was covered with an immersion oil, the peaks shifted to longer wavelengths [7,10]. As a result, the symmetric "red" peak shifted to the infrared part of the spectrum; see the curve 2 of Fig. 2(b) that shows the reflection spectrum for the sample covered with immersion liquid of the refractive index of 1.47. In our experiments, this symmetric plasmon resonance of the double-pillar nano-molecule was excited by the infrared laser of TM polarization and generated strong electromagnetic fields near the sample surface required for the nanometric tweezers operation. It is worth noting that an effective coupling of the infrared laser with plasmon resonances can also be achieved for a single Au dot of a suitable size [3]. We choose the double-pillars in place of single-dots because electron beam lithography provides a better control of the nanocavity "volume" (the region where the excited electromagnetic fields are strong) through the pillar gap (lift-off procedure reliably gave the pillar gap down to 10nm for our geometry) than the dot size (lift-off procedure for dots was reliable down to sizes of 20nm).

Let us describe our experimental results. First, we found that the focused laser beam creates microbubbles in an immersion oil near the described nanostructured substrates. The creation of microbubbles was a controlled and reproducible process characterised by an intensity threshold. The microbubbles were not produced when the laser beam intensity was below the threshold for any time of illumination. When the intensity of the focused laser beam exceeded the threshold, a microbubble was formed near the focus of the beam, see Fig.3. Being created, a bubble grows to a stable state in which the bubble's radius does not change with time provided the leaser power is kept constant. This stable radius can be seen as a plateau (at radius of 14$\mu$m) in Fig. 3(d) which shows the radius of a bubble created at the threshold laser power as a function of time. The stationary state and size of a bubble depended on the laser power with a larger power generating a bigger bubble. Stable bubbles corresponding to a fixed laser power have been used in the following experiments for laser trapping.

Figures 3(a)-(c) demonstrate three different stable microbubbles (indicated by white arrows) created in immersion oil at the threshold for samples with three different pillar diameters. The process of the bubble formation is also shown in movies of the supporting information. It should be noted that the nanostructure of our gold samples played an important role in bubble formation, forasmuch as we did not observe the creation of microbubbles near the surface of a plain layer of gold of the same thickness (80nm) at any laser intensity available in our experiments. We have found that the geometry of nano-molecules (pillar pairs) has a strong influence on the threshold intensity and the size of the bubble created at the threshold. Table 1 gives the characteristics of nanopillars, the threshold power for bubble generation and the diameters of microbubbles created at the threshold for the samples of Fig. 3(a)-(c). We see from Table 1 that samples with pillars of smaller diameters (and hence smaller coupling between pillars) required higher beam intensities to generate microbubbles and resulted in microbubbles of smaller sizes. The threshold intensity depended on the light polarization and was about 1.5 times higher for the TE polarization. Remarkably, the minimal radius of the produced stable bubbles was only about 1μm. The characteristic time of the bubble creation and decay was about $\tau$~0.1s, see the graph of Fig. 3(d) that describes the process of bubble growth to a stable size at the threshold intensity and bubble decay after switching off the laser power.

Second, we found that the created stable microbubbles are efficiently trapped by the laser beam. Trapping and manipulating of the produced bubbles is shown in movies of the supporting information. A microbubble was trapped in an offside position with respect to the focal spot of the beam (usually, the surface of the microbubble was close to the focal spot). A trapped bubble followed the focal spot of the beam focused onto substrate and could be positioned anywhere at a patterned substrate with sub-micrometer precision. (The trapping has been mostly studied for a moving substrate and a fixed laser beam position). When the laser beam had been moved along the patterned substrate in $x$ (or $y$) direction at a constant speed, the trapped bubble followed the focal spot in a continuous manner with a periodic modulation commensurable with the period of the double pillar array. (For small bubbles this modulation can be described as a slow motion of a bubble near the pillar pair and faster "jump-like" motion in between the pairs.) This suggests that a single pair of nanopillars (illuminated by a focused laser beam) could act as an efficient trap for a microbubble. Using a bubble jump as a reference of length, we have evaluated the characteristic trapping distance of *a single pair of nanopillars* to be less than 100nm. The size of a bubble did not change significantly during the bubble trapping and positioning. Figure 4(a) shows the motion of an initially de-trapped bubble toward the trapping position (at the middle of the image), while Figs. 4(b) and (d) show the trapping of bubbles near the focal spot in the middle of the image for the sample substrate moving in $x$ and $y$ directions. The trapping was observed for all sizes of the created stable microbubbles (<100μm). Trapping of larger bubbles is demonstrated in Fig. 4(c). The critical de-trapping speeds were of the order of 100μm/s (Fig. 4(b)), which implies high efficiency of the laser trap, see the discussion below. The de-trapping speed was measured in a conventional way by moving the beam (or substrate) at a constant speed and registering the beam speed at which the bubble de-traps from the laser focus (and hence the trapping force becomes smaller than the Stokes viscous force acting on a bubble in an immersion liquid). When the laser beam was moved away from the area occupied by the array of nano-pillars, microbubbles de-trapped from the laser beam and collapsed with characteristic time $\tau \sim 0.1$s.

We briefly discuss our results. Generation of bubbles near the focus of an intense laser beam is a well-known process usually associated either with thermal absorption [11] or dielectric breakdown in liquid [12]. The thermal absorption mechanism would require light intensity of only $I \approx 10^5$W/cm$^2$ for bubble generation [11] while the dielectric breakdown would require roughly $I \approx 10^{10}$W/cm$^2$ [12]. In our case, intensity of the focused beam was of the order of $I \approx 10^6$-$10^7$W/cm$^2$, which is much larger than that required by the thermal mechanism. In addition, we did not observe bubble formation at the surface of the unpatterned gold strip of the same height. This implies that we can rule out the thermal mechanism as the sole explanation for bubble formation. (Another argument against the thermal mechanism is the dramatic increase in the threshold intensity (from 14mW to 57mW) observed for a relatively modest decrease in the pillars' diameter (from 128nm to 100nm), see Table 1.)

As far as the second mechanism is concerned, the average laser intensity in the focus was three orders of magnitude lower than the intensity required for the dielectric breakdown, which apparently implies that the dielectric breakdown plays no role in bubble formation. However, such conclusion is premature. It is known that the electromagnetic field is greatly enhanced near metallic nanoinclusions [3,7] and, at plasmonic resonances, could be about 10 times larger in the region between the nanopillars than in the incident wave [3,7]. As a result, the peak light intensity in the region between neighbouring nanopillars could be of the order of $I \approx 10^9$W/cm$^2$ in our case. Also, the geometry of the pillar pair restricts motion of the accelerated electrons and could decrease threshold for dielectric breakdown in the presence of a nanostructured substrate. These two facts suggest that we cannot entirely exclude the contribution from dielectric breakdown to bubble formation. Dielectric breakdown would easily explain the dramatic intensity dependence of bubble formation in samples upon gaps between neighbouring nanopillars by correlating the decrease in intensity for smaller gaps $\delta$ (Table 1) with the increase in the field enhancement factor produced by particle near-fields.

There exists, however, a third mechanism for bubble generation, exclusive to nanostructured substrates. Indeed, the gap $\delta$ between nanopillars in the nano-molecule provides an excellent seeding place for bubble formation. The microbubble can form when the peak stress of the electromagnetic field "inside" the nano-molecule prevails the surface tension pressure:

$$q^2 \frac{nP}{c} \approx \frac{2\sigma}{\delta}, \qquad (1)$$

where $q$ is the field enhancement factor due to pillars near fields (we roughly estimate $q \approx Const/\delta$ for the samples of Table 1), $P$ the laser intensity, $n$ the refractive index of the immersion oil, $c$ the speed of light, and $\sigma$ the oil surface tension. In our case both sides of (1) are about $10^5 \mathrm{erg/cm^3}$, which makes possible bubble formation due to stress of the electromagnetic field. Also, from (1) and $q \approx Const/\delta$ it follows that $P/\delta$ should be approximately constant near the threshold if bubbles are to be formed due to electromagnetic stress. It is easy to check that this value is indeed approximately constant ($P/\delta \approx 1.5 \pm 0.3 \, \mathrm{mW/nm}$) for the data shown in Table 1, which supports the stress mechanism. The increase in electromagnetic stress in samples with smaller gaps between pillars (due to larger field enhancement factors) would also qualitatively explain the dramatic increase in size of the developed threshold bubbles shown in Fig. 3(a)-(c). However, at present stage we cannot exclude that all three mechanisms (thermal absorption, dielectric breakdown and electromagnetic stress) may play an important role in bubble formation.

The laser trapping of the generated microbubbles and their nanometric positioning observed in our experiments is less trivial. Indeed, according to the theory of optical trapping, low index particles cannot be trapped by a Gaussian beam propagating in a homogeneous optical medium. The trapping of low index particles requires a special optical trap comprising either a scanning Gaussian beam [13], or a Lageurre-Gaussian beam [14], or a self-focused laser beam [15]. Our work demonstrates that trapping of low index microbubbles can be achieved in *any conventional optical trap (with a Gaussian beam) in the vicinity of nanostructured metal-dielectric substrates*; see also Ref. [11]. The mechanism of microbubble trapping is not entirely clear. Two mechanisms could guarantee the observed effect: i) a thermal mechanism, in which convection induced by laser heating leads to a bubble attraction to the laser beam [11] and ii) an optical mechanism, in which the laser beam excites nanopillar plasmon resonances [3,7] that affect the beam profile near the substrate and (combined with optically induced liquid convection) generate the trapping point near the focus. (It worth noting that the fields produced by vibrating electron plasma in pillar pair are greatly enhanced near plasmon resonances and are concentrated mostly in the region between neighbouring nano-pillars, see Ref. [7].) Our experimental data favours the optical mechanism. Indeed, we observed that the laser beam does not trap microbubbles as soon as the beam (with a bubble) is moved from the nanostructured substrate to an unpatterned gold strip of the same height. This is difficult to explain within the thermal model since the unpatterned strip has stronger thermal absorption than the patterned structure and hence should provide a better thermal trapping. Also, the parameters of the microbubble trapping strongly depend on the geometry of the pillar pair. A relatively small change in the diameter of nanopillars changes dramatically the trapping efficiency (Table I). This is also difficult to explain in terms of the thermal trapping. At the same time, such behaviour should be expected for an optical mechanism of trapping (the geometry of the nanopillar pair strongly affects near-field coupling and excitation of the plasmon resonances in the nanoinclusions). A theory of nanometric optical trapping in the vicinity of a sharp metallic pin was developed in Ref. [3]. We cannot directly compare our results with the theory because of different geometry of metallic inclusions (the theory of laser trapping for our structures is under development). However, it is interesting to note that the calculated light intensity [3] required for the successful operation of nanometric optical tweezers $I \approx 6.5 \times 10^6 \text{W/cm}^2$ is close to intensity used in our experiments for laser trapping and bubble generation. At present, we cannot exclude that both mechanisms (thermal and optical) contribute to the observed trapping of microbubbles. A simple evaluation shows that in our case, on average, there is about one double-pillar nano-molecule in the focal region of the beam. Hence the trap is realised in the vicinity of an illuminated nano-molecule and moves to a next illuminated nano-molecule as the laser beam slowly moves along the sample. This explains the modulated motion of the microbubbles along 2D array of pillar pairs and a size of <100nm of an individual trap observed in our experiments.

Another important parameter of laser tweezers is the trapping efficiency, which is determined as $Q = \frac{F_{tr} c}{nP}$, where $F_{tr}$ is the trapping force. The maximal trapping force can be evaluated from the de-trapping speeds $V$ of stable bubbles with the help of Stokes' law modified by the presence of a substrate $F_{tr} = K \cdot 6\pi\eta r V$ [16], where $K$ is the correction coefficient due to proximity to the substrate ($K \approx 2.5$ in our geometry), $\eta \approx 140$ g/(ms) is the immersion oil viscosity, $r$ is the bubble radius. The de-trapping speeds have been measured at the laser power of $P \approx 60$mW and the trapping efficiencies are given in the Table 1. Remarkably, the sample of Fig. 3(a) yields giant trapping efficiency of $Q \approx 45$ (for the microbubbles of the size of 18μm). For the smaller bubbles of $r \approx 3$μm (the sample of Fig. 3(c) and Fig. 4), the de-trapping speed was about $V \approx 70$μm/s. This implies a very large trapping force $F_{tr} \approx 1.4$nN and still very large trapping efficiency $Q \approx 5$. These values observed for nanometric laser trapping of microbubbles near the nanostructured substrates are two-three orders of magnitudes higher than those observed for conventional optical tweezers.

The proposed scheme of nanometric laser tweezers holds a big promise for a whole set of different applications. First, the studied nanostructured substrates can be readily used in any standard laser tweezers and can improve their basic characteristics by increasing the trapping force, the trapping efficiency, and providing sub-micrometer level of particle positioning. Second, the described nanometric tweezers can be readily combined with interferometric or dark field techniques for selective chemical [17] and bio [18, 19] analyses. Third, the proposed arrays of metallic nano-pillars can play a role of surface enhanced Raman substrates [10] in Raman tweezers [20], where focused laser beams are used for object positioning (manipulating) as well as its analysis. From the other hand, generation of microbubbles near a nanostructured substrate may also find its applications in science and technology. These include bubble valves, switches or manipulators, see Ref. [21], tuneable microfluidic optical fibre gratings [22], miniature bubble lenses [23], etc.

In conclusion, we have demonstrated nanometric laser tweezers based on an array of metallic nano-molecules consisting of pairs of gold nanoparticles. Using low intensity laser beams focused on a nanostructured substrate consisting of an array of gold nanopillars, we have created, trapped and manipulated microbubbles in an immersion liquid. The mechanism of trapping of microbubbles near the surface of nanostructured substrates is still to be clarified. The nanometric laser tweezers showed a sub-micron precision of microbubble positioning and a giant trapping efficiency ($Q \sim 50$). We believe that the proposed nano-structured substrates could be a routine addition to conventional optical tweezers.

**Supporting Information Available**: Movies of microbubble creation, manipulation and trapping. This material is available free of charge via the Internet at http://pubs.acs.org.

**Figure Captions.**

Fig. 1. Schematics of laser tweezers installation.

Fig. 2. Gold nanopillar arrays. (a) Scanning electron micrograph of the sample. (b) The TM reflection spectra of the nanopillar array shown in (a) measured in air, 1, and for a sample covered by an immersion oil, 2. The inset shows the micrograph of a sample viewed at an angle.

Fig. 3. Microbubble formation. Microbubbles generated near the surface of the nanostructured substrates at threshold intensity for samples with three different diameters of nanopillars (a) D=128nm (b) D=115nm (c) D=100nm. All arrays have a period of 400nm, the pillar separation of 140nm and appear to be greenish on the images. The white arrows indicate microbubbles, the white scale at the right-bottom corner of the images is 10μm. (c) The dynamics of a bubble radius at the threshold intensity and after switching the laser power off. The sample is the same as in Fig. 3(a).

Fig. 4. Microbubble trapping (a) and manipulating (b)-(d). The greenish area is the patterned part of the glass substrate with nanofabricated arrays of nano-pillars. The focus of the beam (the trapping point) is in the centre of the images. The times elapsed between frames are (a) 0.03s, (b) 0.12s (c) 0.3s (d) 0.18s. The sample is the same as in Fig. 3(c). The white dotted arrows indicate a displacement of a substrate or a bubble. The white scale at the right-bottom corner of the images is 10μm.

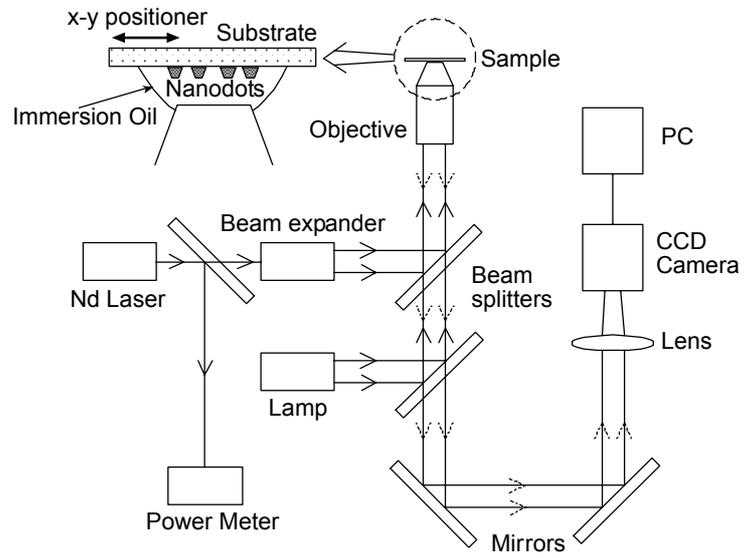

Fig. 1

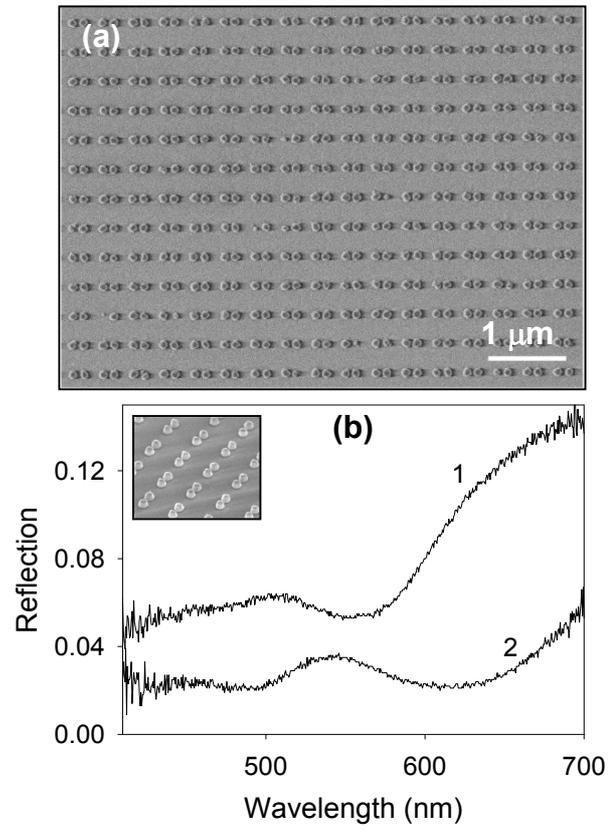

Fig. 2

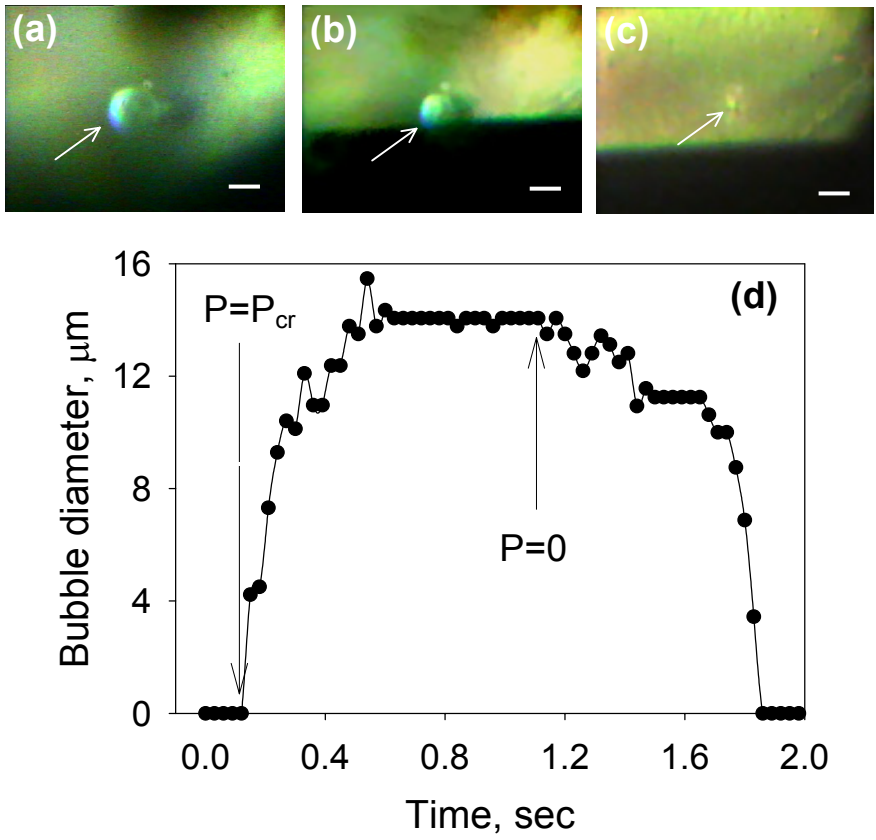

Fig. 3

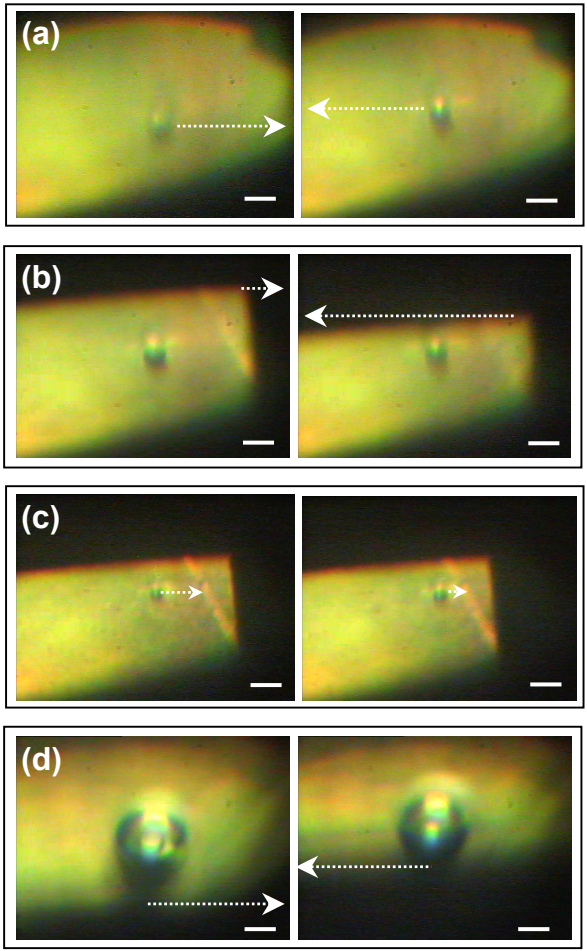

Fig. 4

**Table 1**
Parameters of bubble generation and trapping.

| Samples of (a=400nm, s=140nm) | Diameter of pillars (nm) | Threshold power for bubble generation (mW) | Diameter of a bubble at threshold (μm) | De-trapping speed at 60mW (μm/s) | Trapping efficiency Q |
|---|---|---|---|---|---|
| Fig. 3(a) | 128 | 14 | 14 | 70 | 45 |
| Fig. 3(b) | 115 | 45 | 10 | 50 | 20 |
| Fig. 3(c) | 100 | 57 | 2 | 70 | 5 |